\documentclass{aa520}  % for me
\usepackage{graphicx}
\begin{document}

\title{Companions of old brown dwarfs, and very low mass stars
   \thanks{Based on observations obtained at the European 
           Southern Observatory, Paranal with the VLT UT2 (KUEYEN). 
           ESO Proposals 68.C-0063, 67.C-0160.}}

   \author{Eike W. Guenther \inst{1} \and G\"unther
          Wuchterl\inst{2}}

   \offprints{E. Guenther}

   \institute{Thueringer Landessternwarte Tautenburg, 
              Sternwarte 5 
              07778 Tautenburg, Germany \\
             \email{guenther@tls-tautenburg.de}
         \and
             Max-Planck-Institut f\"ur extraterrestrische Physik (MPE)
             Giessenbachstr., 85741 Garching, Germany \\ 
             \email{wuchterl@mpe.mpg.de}
             }

   \date{Received September 6, 2002; Accepted 24 January 2003}

\abstract{Up to now, most planet search projects have concentrated on F
to K stars. In order to considerably widen the view, we have stated a
survey for planets of old, nearby brown dwarfs and very low mass
stars. Using UVES, we have observed 26 brown dwarfs and very low mass
stars. These objects are quite inactive and are thus highly suitable for
such a project.  Two objects were found to be spectroscopic
binaries. Another object shows significant radial velocity variations.
From our measurements, we conclude that this object either has a
planetary-mass companion, or the variations are caused by surface
features. Within the errors of the measurements, the remaining objects
are constant in radial velocity.  While it is impossible to strictly
exclude an orbiting planet from sparsely sampled RV data, we conclude
that it is unlikely that these objects are orbited by massive planets
with periods of 40\,days or less.  }

\maketitle
%
%________________________________________________________________

\section{Introduction}

Precise measurements of the radial velocity (RV) of stars have lead to
the discovery of more than 100 extrasolar planets.  From these surveys
it is concluded that at least 3\% of G to K stars harbor giant planets
(Queloz, Santos, Mayor \cite{queloz02}). The true fraction is certainly
higher than that, since the mass-function for planets rises steeply
towards smaller masses, and planets in long-period orbits are difficult
to detect in RV-surveys.  Most of the efforts for detecting extrasolar
planets have hitherto been concentrated on old, solar-like F-K stars.
The next logical step is to widen the view in order to find out which
properties of the central star influences the presence or absence of
planets. An important parameter is of course the mass of the central
star. Since detecting planets in orbit around massive stars is rather
difficult, the surveys should be widened to include objects of much
smaller mass.  A few surveys for planets of M-stars are now
underway. Marcy et al.  (\cite{marcy98}) have studied 24 M-dwarfs for 4
years, and found a planet with $M\,sin\,i=2.1 M_{\rm J}$ in orbit around
the M4 dwarf Gl~876 (0.32\,$M_{\sun}$). Delfosse et
al. (\cite{delfosse98}; \cite{delfosse99}) have surveyed all known
M-dwarfs within 9~pc, north of -17 declination, and brighter than 15th
mag in V. In total, the sample contained 136 stars. The RV-measurements
had an accuracy between 10 and 70 m/s, sufficient for detecting giant
planets. Apart from the planet of Gl~876 which was found independently,
no other planet was found.  Nearby low-mass objects are also suitable
for searching long-period planets astrometrically. However, theses
surveys yieled upper limits. For example, using the HST Fine
Guidance Sensors, Benedict et al. (\cite{benedict98}) could exclude
planets more massive than 0.3$M_{\rm J}$ and with periods of $\sim$ 600
days in orbit around Proxima Centauri and Barnard's Star (0.10 and 0.12
$M_{\sun}$). An upper limit of the frequency of planets was also derived
from microlensing data.  Gaudi et al. (\cite{gaudi02}) conclude from an
analysis of microlensing data that less than 33\% of the M-dwarfs in the
Galactic bulge have companions of 1.0$M_{\rm J}$ between 1.5 and 4
AU. Up to now, all surveys of M-dwarfs have focused on stars with masses
of 0.3 $M_{\sun}$, or so. This is only a factor three lower than the
G-dwarf surveys. In order to really find out what the influence of the
mass of central star is, it is thus necessary to go to objects of much
lower mass. In here we report on a survey of very-low-mass stars (VLMSs)
and brown dwarfs (BDs). All of them have masses of 0.1 $M_{\sun}$ or
less. Surveys of visual companions of BDs already have identified a
number of BD-BD companions (Mart\'\i n et al.\cite {martin00}, Lane et
al. \cite{lane01}, Kenworthy et al. \cite{kenworthy01}, Reid et
al. \cite{reid01}, Close et al. \cite{close02a}, Potter et
al. \cite{potter02}, Goto et al. \cite{goto02}). Additionally, a
spectroscopic BD-BD companion has also been found (Basri \& Mart\'\i n
\cite{basri99}). This binary consists of two BDs with masses of 0.06 to
0.07 $M_{\sun}$. The orbital period is 5.8 days, the eccentricity
$0.4\pm0.05$.  While the aim of this project is to find out whether BDs
and VLMs have planets, we will briefly review the arguments for and
against the presence of giant planets in orbit around such low-mass
objects first.

The first argument for the possible presence of planets in orbit around
VLMSs and BDs is that these objects have accretion disks when they
are young.  This is the result of the work by Comer\'on et
al. (\cite{cameron00}), who find an excess of the emission at 6.7 $\mu
m$ for 4 out of 13 VLMSs and BDs in the Chamaeleon\,I star forming
region. Muench et al. (\cite{muench01}) concludes from the JHK
colour-colour diagram of the Trapezium Cluster in Orion that the
disk-frequency of low-mass members of the cluster is 50~\%, or higher.
An excess of infrared emission of low-mass objects was also observed
with ISOCAM in the Chameleon I dark cloud (Persi et
al. \cite{persi00}). Young, low mass objects apparently not only have a
passive disks but also show signs of accretion, and outflow activity
like in T\,Tauri stars (Muzerolle et al.  \cite{muzerolle00},
Mart\'\i n et al. \cite{martin01b}, Fern\'andez \& Comer\'on
\cite{fernandez01}).  Accretion from a disk seems to be present even for
objects with a mass of only 8-12 $M_{\rm J}$ (Testi et
al. \cite{testti02}). We thus can conclude that many young brown dwarfs
have disks but are theses disks massive enough to form {\em giant}
planets?

Disks of T\,Tauri stars have typically about 3\% of the mass of the
central object. If this is also true for BDs, the total mass of a a
typical BD disk would be only a few $M_{\rm J}$. However, observation
Cha\,$H\alpha$\,2 at 9.8 and 11.9 $\mu m$ show that the disks of BDs are
not just scaled down versions of T~Tauri star accretion disks (Apai et
al. \cite{apai02}). We thus have to wait until mm-wavelength
observations become available, to find out what the true masses of the
disks of young BDs are. On the other hand, a small mass of the central
object may help the formation of giant planets, because the small mass
of the central object will result in relatively larger Roche-Lobes and
larger feeding zones at typical planetary masses.

Giant planets may form either by a gravitational instability of the
disk, or by accreting disk-gas onto a solid core of a few earth-masses
(see review in Lin et al. \cite{lin00}; Ward \& Hahn \cite{ward00};
Wuchterl, Guillot, Lissauer \cite{wuchterl00}). Before the discovery of
the first extrasolar planets, it was generally believed that most giant
planets would be found close to the so called snow-radius. The idea was
that since the growth-rates of planete\-simals increases at the snow
radius because the amount of condensed elements increases, giant planets
would also be found there (Lunine \& Stevenson \cite{Lunine88}). At
distances much larger than the snow-radius the formation of giant planet
would then be very slow.  After the discovery of massive planets close
to the stars, a number of scenarios for their formation were
proposed. Wuchterl (\cite{wuchterl93}) argued that the accretion of
massive envelopes of giant planets points to a favorable zone inside of
0.1 AU. Another idea was that massive planets form indeed close to the
snow-radius but then migrate inwards (Trilling, et
al. \cite{trilling98}).  According to Armitage and Bonnell
(\cite{armitage02}), orbital migration also causes the so called
BD-desert. These authors argue that orbital migration depends on the
ratio between the disk-mass and the mass of the planet, and since the
mass of the accretion disk of a 1.0$M\sun$ star corresponds to the mass
of a BD, the orbital migration would become so strong that such an
object would fall into the star.  If this is true, it would imply that
giant planets of BDs could not exist, because the disk of a BD seems to
be about 1.0$M_J$.

The tidal interaction between the disk and the planet would always pull
the planet {\em inward}, and we thus expect to find planets at the
snow-radius, or at {\em smaller} distances from the star. The position
of the snow-line has recently been calculated by Sasselov and Lecar
(Sasselov \& Lecar \cite{sasselov00}). The snow-radius is at 0.6 AU for
solar-mass star and a passive disk. Using the formula published by
Sasselov and Lecar (\cite{sasselov00}), we find that the snow-line
corresponds to orbital periods between 20, and 40 days in the case of
VLMSs and BDs. It is interesting to note that the surface density
at the snow-radius is roughly the same for the disk of a solar-like
star, and the disk of a BD.  Giant planets thus are expected to have
periods of 40\,days or less if they exist.  A one $M_{\rm J}$-planet in
a circular orbit would induce RV-variations with a amplitude of
750\,m/s. As discussed in Desidera (\cite{desidera99}), a planet
orbiting a BD at very small distances will be tidally disrupted.  This
gives us a lower limit for the orbital period. For a one $M_{\rm J}$
planet orbiting an old 0.070 $M_{\sun}$ BD, this limit corresponds to an
orbital period of 15 hours. The RV-variations induced by such a planet
would be 2800\,m/s. In summary, if BDs have giant planets, we expect
them to have orbital periods between about half a day and 40 days. The
aim of this work is to search for such objects.

%__________________________________________________________________

\section{Observations}

Suitable targets for our project are BDs and very late-type stars close
to the sun. The spectroscopic observations were carried out with the
Uv-Visual Echelle Spectrograph (UVES) on the VLT Unit telescope 2
(KUEYEN) in service mode. Because the objects are extremely red, we use
the setting which covers simultaneously the wavelength regions from 6670
to 8545 \AA , and 8640 to 10400 \AA . This wavelength regions also
contains the telluric bands between 6860 to 6930 \AA , and 7600 to 7700
\AA \, which we used as a secondary wavelength reference. Before the
introduction of the iodine-cell, the telluric lines were often used
for similar purposes in the past. The accuracy and the limitations of
the method thus have been investigated thoroughly: Using the Kitt Peak
1-m Vacuum Fourier Transform Spectrometer, Balthasar et
al. (\cite{balthasar82}) have studied the line-shift of eleven telluric
$O_2$-line in detail. The shifts were found to be 15 m/s, or less. These
results agree well with results of Smith (\cite{smith82}) who find that
RV-measurements of stars have typical errors of 8 m/s if the telluric
lines are used as wavelength reference.  Caccin et al. (\cite{caccin85})
confirm these results and find that the shifts are caused by the wind
in the earth atmosphere.

Standard IRAF routines of the Echelle package were used to
flat-field and wavelength calibrate the spectra using frames taken with
the standard flat-field and Thorium Argon lamps. Since the objects are
relatively faint, the extraction was done in a different manner than
usual. In the first step we extracted each order as a two dimensional
spectrum, then subtracted the sky-background in the two-dimensional
images, and finally extracted the spectrum of the object. No rebinning
was done in order to achieve the highest possible accuracy for the
RV-measurements. Except for the sky-subtraction, the data was thus
reduced in the same way as in our previous UVES observations (Joergens
\& Guenther \cite{joergens01}). In that run we took always two spectra
of the same object directly after each other in order to 
remove cosmic rays. Because we always took two spectra per night, we could
calculate the error from the scatter of the measurements. The result was
that the errors are 90 m/s for the brighter sources.  However, the
objects discussed in this paper are fainter, and the S/N-ratio is
significantly lower than in our previous paper. Most of the
RV-measurements are thus limited by the S/N-ratio to 200-300~m/s.
Table~\ref{obslog} gives an overview of the observations. 

%__________________________________________________ One column table
   \begin{table} \caption[]{Observing log} \label{obslog}
     $$ 
         \begin{array}{lcll}
            \hline
            \noalign{\smallskip}
object & No.\,of  & spectral & average \\
       & spectra  & type   & S/N^d \\
            \noalign{\smallskip}
            \hline
            \noalign{\smallskip}
2MASSW\,J0832045-012835 & 2 & L1.5 & 17, 25 \\
2MASSW\,J0952219-192431^a & 1 & M7   & 71 \\
2MASSW\,J1237270-211748 & 3 & M6   & 34, 33, 26 \\
2MASSW\,J2013510-313651 & 3 & M6   & 33, 37, 39 \\
2MASSW\,J2049197-194432 & 3 & M7.5 & 22, 25, 24 \\
2MASSW\,J2052086-231809 & 3 & M6.5 & 40, 43, 38 \\
2MASSW\,J2113029-100941^c & 3 & M6   & 30, 33, 30 \\
2MASSW\,J2135146-315345 & 3 & M6   & 32, 26, 40 \\
2MASSW\,J2147446-264406 & 3 & M7.5 & 20, 22, 22 \\
2MASSW\,J2202112-110946 & 3 & M6.5 & 31, 34, 34 \\
2MASSW\,J2206228-204705^b & 3 & M8 & 27, 39, 37 \\ 
2MASSW\,J2306292-050227 & 2 & M7.5 & 13, 54 \\
BRI\,B0021-0214         & 3 & M9.5 & 31, 25, 29 \\
BRI\,B0246-1703         & 2 & M8   & 26, 27 \\
BRI\,B1104-1227         & 3 & M6.5 & 59, 65, 60 \\
BRI\,B1507-0229         & 3 & M6   & 27, 16, 28 \\
Denis-P\,J0021.0-4244   & 2 & M9.5 & 26, 26 \\
LHS\,2065               & 3 & M9   & 42, 62, 60 \\
LHS\,2397               & 4 & M8   & 36, 38, 38, 37 \\
LHS\,292^c              & 3 & M6.5 & 187, 174, 185 \\
LHS\,3566               & 2 & M8.5 & 64, 61 \\
UScoCTIO-055            & 2 & M5.5 & 27, 44 \\
UScoCTIO-075            & 3 & M6   & 24, 35, 28 \\
UScoCTIO-085            & 3 & M6   & 27, 19, 28 \\
UScoCTIO-100            & 3 & M7   & 33, 31, 26 \\
LP\,944-20              & 6 & M9   & 51, 68, 67, \\ 
LP\,944-20              &   &      & 65, 62, 67 \\
            \noalign{\smallskip}
            \hline
         \end{array}
     $$ 
\begin{list}{}{}
\item[$^{\mathrm{a}}$] It was recently discovered that this
                       object is a spectroscopic binary 
                       (Reid et al. (\cite{reid02}).
\item[$^{\mathrm{b}}$] This object recently turned out to be a 
                       visual binary with a separation
                       of $4.1\pm1.1$~AU (Close et al. \cite{close02b}). 
\item[$^{\mathrm{c}}$] In this work we show that this object
                       is a spectroscopic binary.
\item[$^{\mathrm{d}}$] The S/N-ratios of the spectra are derived 
                       from the number of photons detected in
                       the wavelength-region between 
                       6700 and 8500 \AA , and taking the 
                       read-out-noise into account.
\end{list}
   \end{table}

\section{Limits of the masses of companions}

%_____________________________________________________________
%                 A figure as large as the width of the column
%-------------------------------------------------------------
   \begin{figure*}
   \centering
   \includegraphics[width=\textwidth]{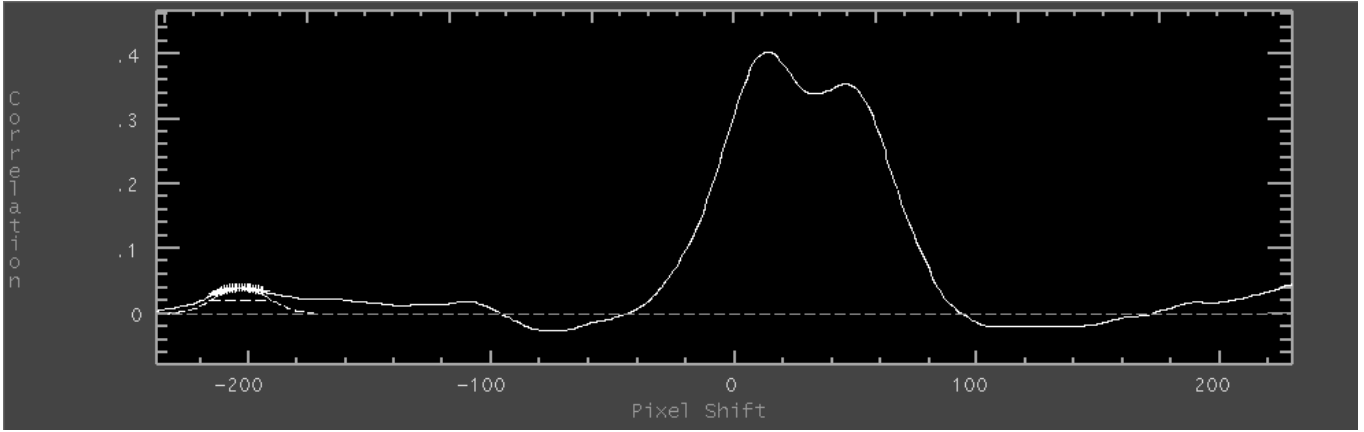}
      \caption{Cross-correlation function of 2MASSW\,J2113029-100941, 
               demonstrating that it is a double-line spectroscopic 
               binary.}
         \label{2mwj2113b}
   \end{figure*}
%
%_____________________________________________________________

All except one of the 25 objects were observed at least twice.  The
results are listed in Table~\ref{rv}. The first column gives the name of
the star, the second the heliocentric RV of the first spectrum. Listed
in the third column is the time-difference between the first and the
second spectrum. The fourth column gives the difference in RV between
the two spectra as measured with the cross-correlation method. The last
two columns are the same for the first and third spectrum. For
determining the heliocentric RV, we had to measure the position of
individual spectral lines, which is of course far less accurate
than using the cross-correlation. The error of these values are
1.3 km/s. In the case of LHS\,2397 four spectra where taken, and in the
case of LP\,944-20 six. Listed in Table~\ref{rv} are always the time and
RV differences to the first spectrum.  The errors of the difference of
RVs given in Table~\ref{rv} are derived from the errors of the
determination of the position of the telluric lines, and the
photospheric lines of the two spectra. The errors of the difference of
two RVs are thus larger than the determination of the RV for each of the
objects. Nevertheless, since we used the cross-correlation, the errors
are still much smaller than the error of the heliocentric
RV. 2MASSW\,J2113029-100941, LHS\,292 and 2MASSW\,J0952219-192431 will
be discussed in the next section, and LP\,944-20 will be discussed
together with LHS\,2065 in section~5. In all other case the $\Delta\,RV$
is less than 3~$\Sigma$.

Deriving a unique upper limit for the masses of possible companions from
the $\Delta$~RV-values is not possible, because the $\Delta$~RV-values
depend on:

\begin{itemize}
\item the eccentricity of the orbit and if it is eccentric also 
      on the node angle,
\item the inclination of the orbit,
\item the orbital period of the planet, and
\item the orbital phase when the first observation was taken.
\end{itemize}

In order to avoid giving large tables for all possible combination of
parameters, we prefer to discuss just a typical case. That is, we
calculate the $\Delta\,RV$-values for all phases, given the
time-difference of the measurement, and mass of the central object.  In
order to limit the number of free parameters, we assume a round orbit.
Given in the last column in Table~\ref{rv} is then the mass of planet
($m\,sin\,i$) that would have been detected with a probability of $\geq
50$\%.  As we outlined in the introduction, we expect that giant planets
have periods of 40\,days or less. We thus compute the upper limit for an
orbital period of 40\,days. However, that introduces the problem that
the upper limit could not be derived for UScoCTIO-055, and UScoCTIO-075.
In these two cases we thus use a period of 30\,days.  Except for
BRI\,B1507-0229, the upper limits are a few $M_{\rm J}$.  However,
because of all these restrictions, the upper limits given in
Table~\ref{rv} should not be taken as strict upper limits but as
guidelines for the typical mass of companions that can be excluded. With
just two or three RV-measurements there is always an orbit where
$\Delta_{1,2,3}\,RV\,=\,0.$ but $m_{companion}\,sin\,i$ is large. It is
just that it is not likely that all planets of all BDs are in such
orbits. We thus safely conclude that massive planets are not common for
VLMSs and BDs. Possibly none of these objects is orbited by a massive
planet with a period of 40\,days.

Another interesting question is whether we can exclude a planet like
the one of 51\,Peg. The planet of 51\,Peg has an $m\,sin\,i$ of 0.47
$M_{\rm J}$, and orbits the star with a period of 4.2\,days. Although
such a planet orbiting a 0.08 $M_{\sun}$ object would induce
RV-variations with an amplitude of about 700 m/s, the sampling of the
data allows to exclude such an object only in 10 of 23 objects
(2MASSW\, J1237270-211748, 2MASSW\, J2049197-194432,
2MASSW\, J2113029-100941, 2MASSW\, J2135146-315345, BRI\, B0021-0214,
Denis-P\, J0021.0-4244, 2MASSW\, J2013510-313651, LHS\, 2065, 
LHS\,3566, UScoCTIO-085).

%__________________________________________________ One column table
   \begin{table*} \caption[]{RV-measurements. Because the 
    cross-correlation method is used and the absolute
    velocity of the templates is unknown, we give here
    only the differences in velocity between the first,
    and the other spectra.} \label{rv}
     $$ 
         \begin{array}{lrrrrrc}
            \hline
            \noalign{\smallskip}
object & RV^d & \Delta t & \Delta RV & \Delta t & \Delta RV & upper limits \\
       & & 1-2      & 1-2 & 1-3 & 1-3                  & for\,m\sin\,i \\
       & [km/s] &  [days]  & [km/s]    & [days]   & [km/s]    & [M_{\rm J}] ^a\\
            \noalign{\smallskip}
            \hline
            \noalign{\smallskip}
2MASSW\,J0832045-012835 & 20.0  & 55.9 &  1.30\pm1.23 &      &              & 
4.0 \\ 
2MASSW\,J1237270-211748 & -8.7  & 35.1 & -0.37\pm0.14 & 71.6 & 0.14\pm0.31  & 
0.9 \\
2MASSW\,J2013510-313651 & -21.3 & 61.8 &  0.19\pm0.30 & 71.6 & 0.33\pm0.30  & 
0.6 \\
2MASSW\,J2049197-194432 & -41.5 & 8.8  &  0.08\pm0.30 & 27.8 &  0.26\pm0.29 & 
0.4 \\ 
2MASSW\,J2052086-231809 &  34.4 & 8.8  & -0.51\pm0.33 & 27.8 & -0.57\pm0.31 & 
2.2 \\ 
2MASSW\,J2113029-100941 &  -6.4 & 8.8  & -0.19\pm0.52 & 27.8 &  0.54\pm0.51 & 
0.8 \\
2MASSW\,J2135146-315345 &  17.7 & 62.9 &  0.75\pm0.33 & 72.7 &  0.09\pm0.34 & 
0.5 \\
2MASSW\,J2147446-264406 &  31.2 & 63.9 &  1.07\pm0.52 & 75.7 &  0.99\pm0.41 & 
2.9 \\
2MASSW\,J2202112-110946 &  -9.4 & 8.9  & -0.75\pm0.68 & 27.8 & -0.99\pm0.74 & 
3.2 \\
2MASSW\,J2206228-204705^c & 11.2 & 8.9  & -0.65\pm0.44 & 27.8 & -0.53\pm0.54 & 
1.7 \\ 
2MASSW\,J2306292-050227 & - & 9.8  & useless &      &                   &  - \\
BRI\,B0021-0214         & 16.0 & 28.0 & -0.18\pm0.20 & 41.8 & useless      & 0.6 
\\
BRI\,B0246-1703         &  5.6 & 34.9 &  0.58\pm0.13 &      &              & 4.2 
\\
BRI\,B1104-1227         &  9.6 & 31.8 &  0.46\pm0.17 & 56.8 & useless      & 2.2 
\\
BRI\,B1507-0229         & -37.7 & 10.1 & -2.14\pm2.90 & 27.1 & -2.50\pm2.8  & 
8.4 \\
Denis-P\,J0021.0-4244   &  3.7 & 43.9 & -0.30\pm0.27 &      &              & 2.3 
\\
LHS\,2065               &  7.1 & 8.9  & -1.13\pm0.85 & 31.8 & -0.26\pm0.25 & 1.1 
\\
LHS\,2397               & 33.1 & 29.8 & -2.19\pm0.75 & 38.7 & -0.25\pm0.99 & \\
LHS\,2397               &      & 51.7 & -0.95\pm0.88 &      &              & 3.4 
\\
LHS\,292                &  1.7 & 40.0 &  9.38\pm0.55 & 53.7 & -0.10\pm0.45 & 
^f\\ 
LHS\,3566               & -21.5 & 71.7 &  0.20\pm0.27 &      &              & 
1.1 \\
UScoCTIO-055    & -5.4 & 20.0 &  0.75\pm0.27 &      &              & 2.5^b \\
UScoCTIO-075    & -5.3 & 20.0 & -0.74\pm1.47 &      &              & 2.5^b \\
UScoCTIO-085    & -21.3 & 17.0 & -0.35\pm0.50 & 27.0 & 0.26\pm0.40  & 0.9\\
UScoCTIO-100    & -3.6 & 19.0 &  2.16\pm1.50 & 25.9 & 0.35\pm1.50  & 1.0 \\
LP\,944-20^e    &  8.9 & 23.9 & -0.45\pm0.14 & 24.9 & -1.24\pm1.44  & \\
LP\,944-20      &      & 35.0 &  0.37\pm0.09 & 47.0 & -1.06\pm0.22 & \\
LP\,944-20      &      & 48.1 & -0.52\pm0.76 & & & \\
            \noalign{\smallskip}
            \hline
         \end{array}
     $$ 
\begin{list}{}{}
\item[$^{\mathrm{a}}$] Upper limits of the mass of the companion
                       assuming a planet in a circular orbit
                       with a period of 40\,days.
\item[$^{\mathrm{b}}$] With $\Delta\,t$ of 20 days it makes no sense to
                       compute an upper limit for a 40\,day period,
                       we thus use a 30\,day period. 
\item[$^{\mathrm{c}}$] This object recently turned out to be a 
                       visual binary with a separation of $4.1\pm1.1$~AU 
                       (Close et al. \cite{close02b}), which might have
                       influenced the RV determination of this object. 
                       Reid et al. (\cite{reid02}) gives $8\pm2.0$~km/s as 
                       RV of $H\alpha$, and $16.3\pm2.7$~km/s for the 
                       photospheric lines. We find $10.8\pm1.3$~km/s which 
                       is in the middle of the two.
\item[$^{\mathrm{d}}$] Absolute RV of the first spectrum. Because
                       this measurements where made by measuring 
                       individual lines, the accuracy is only 1.3 km/s.
\item[$^{\mathrm{e}}$] Shown in the case of LP\,944-20 are always the
                       time, and velocity difference between the first, and 
                       the other five spectra. No upper limit is
                       given, because the object shows RV-variations.
\item[$^{\mathrm{f}}$] This object is presumably a spectroscopic binary. 
\end{list}
   \end{table*}

\section{The spectroscopic binaries LHS\,292, 2MASSW\, J2113-1009, 
and 2MAASSW\, J0952-1924}

We took three spectra of the M6V-star 2MASSW\,J2113029-100941 at
MJD\,52106.317, 52115.107, and 52134.091. The first spectrum shows a
nice Gaussian cross-correlation function. In the second spectrum the
cross-correlation function is 4 km/s broader than in the first. The
third spectrum then shows two peaks of about the same height and
separated by 13~km/s (see Fig.~\ref{2mwj2113b}). We thus conclude that
2MASSW\,J2113029-100941 is a double-line spectroscopic binary. Assuming
an age of the order of 5 Gyrs and using the evolutionary tracks from
Chabrier et al. (\cite{chabrier00}) the mass for both components are
between 0.08 and 0.09 $M_{\sun}$. The distance is about 10 to
15\,pc. From the separation of the peaks in the third spectrum we
conclude that the orbital period has to be less than 3 years if the
eccentricity is 0.4 or less.

As can be seen from the Table~\ref{rv}, LHS\,292 shows significant
RV-variations. The first and the second spectrum of LHS\,292 have about
the same RV, and the cross-correlation peak has the same width. In
contrast to this, the cross-correlation peak of the last spectrum is
10~km/s broader, asymmetric and the centre of mass is shifted by about
9~km/s in respect to the two other spectra. The lines are also
shallower in the last spectrum than in the other two. The best
explanation is that LHS\,292 is also a double-line
spectroscopic binary where one component is fainter than the other one.
In this respect it is interesting to note that LHS\,292 is known to show
flare activity (Mullan \& Mac Donald \cite{mullan02}) despite its old
age.

Reid et al. (\cite{reid02}) recently discovered that 2MASSW\,
J0952219-192431 is a double-line spectroscopic binary. We took just one
spectrum of this star which shows only a single peak.

\section{The two active objects LP\,944-20 and LHS\,2065}

%_____________________________________________________________
%                 A figure as large as the width of the column
%-------------------------------------------------------------
   \begin{figure*}
   \centering
   \includegraphics[width=\textwidth]{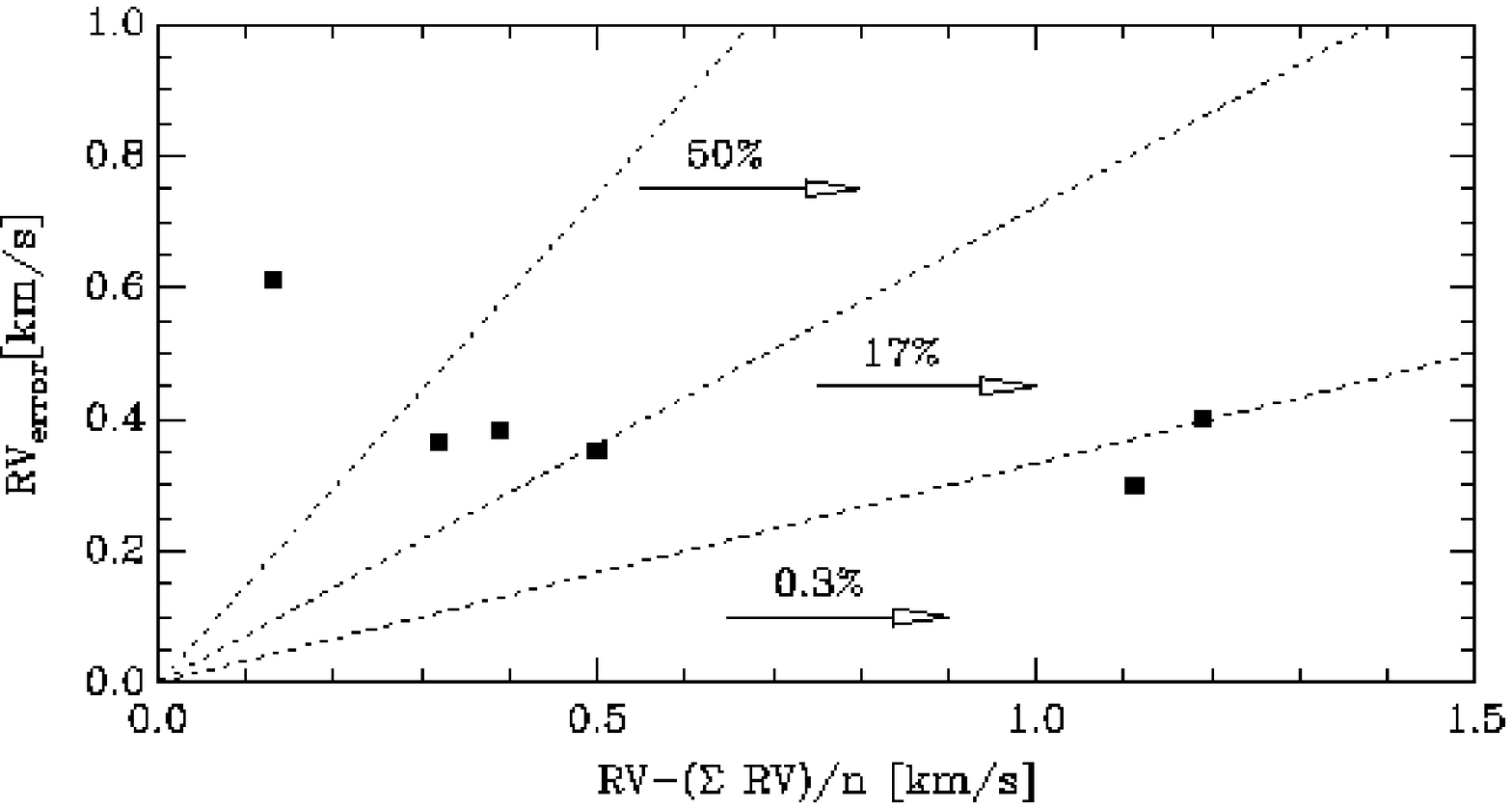}
      \caption{Does LP\,944-20 show significant RV-variations?
               Shown are the RV-measurements (mean value 
               subtracted), versus the error of the RV.
               If LP\,944-20 would be constant, only three
               of the points should be right of the 50\%-line,
               and only one right of the 17\%-line. However,
               we find that 5 out of 6 points are right of the
               first, and 3 are right of the 17\%-line. Two are at, 
               or even right the 0.3\% line. The probability that 
               this is a chance coincidence is 0.0001. We thus conclude 
               that the RV of LP\,944-20 is variable.
               }
         \label{lp944_rv}
   \end{figure*}
%
%_____________________________________________________________

LP\,944-20 (=BRI\,B0337-3535) is an isolated, non-accreting, BD
identified through its Li abundance and low luminosity (Mart\'\i n \&
Bouy \cite{martin02}).  The mass of LP\,944-20 is estimated to be
between 0.056 and 0.064 $M_{\sun}$, and its age is between 475 and 650
Myrs (Tinney \& Reid \cite{tinney98b}). Recent flux-measurements at 5,
9.8 and 11.9 $m\mu$ do not show any excess emission (Apai et
al. \cite{apai02}), which implies that there is no disc.  As pointed out
by Tinney \& Reid (\cite{tinney98b}), LP\,944-20 is one of those very
late type objects that are quite inactive despite their rapid rotation
($v\,sin\,i$ of 28 km/s).  The same conclusion can be drawn from our
spectra. We also derive that the object is quite rapidly
rotating($v\,sin\,i=32\pm4$ km/s) but no infrared CaII emission lines
($\lambda$ 8498 \AA , 8542 \AA , 8662 \AA ) are seen. The X-ray emission
in quiescence is so small that it has not been detected yet (Rutledge et
al. \cite{rutledge00}; Mart\'\i n \& Bouy \cite{martin02}).  On the
other hand, the object is a non-thermal radio source (Berger et
al. \cite{berger01}), and a flare has been detected in the X-rays
(Rutledge et al. \cite{rutledge00}).  Tinney \& Tolley (\cite{tinney99})
have detected small brightness variations of 0.04 mag. These are
equivalent to $T_{\rm eff}$-variations of 20\,K over the entire visible
disc, or 400\,K over 5\% of the disc. Using the optical monitor of XMM,
Mart\'\i n \& Bouy (\cite{martin02}) also find indication for a possible
variability (standard deviation $\leq 0.08$ mag).

We obtained six spectra of this object. The average heliocentric
velocity is $8.3\pm1.3$~km/s which agrees well with $10.0\pm2.0$~km/s
given by Tinney (\cite{tinney98a}).  The $\Delta\,RV$-values derived are
given in Table~\ref{rv}. At first glance the object seems to be variable
but a more detailed analysis is needed. The crucial question is whether
the RV-variations are still consistent with the errors, or larger
than that. As a first step we carefully derive the error of the
measurements by cross-correlating the spectra of LP\,944-20 with a
number of different templates. We then derive the total error for each
data-point from the scatter of these values, taking also the error of
the telluric correction into account. Shown in Fig.~\ref{lp944_rv} on
the X-axis the RV-variations (absolute-value of the difference between
the measured RV and the average RV), and the error of each measurement
on the Y-axis.  The leftmost dashed line in Fig.~\ref{lp944_rv} shows
the dividing line for 50\%. That is, if the object is not variable, 50\%
of the data-points would be left of this line, and 50\% right of
it.  However, 5 out of 6 measurements are right of this line. This
indicates that the scatter of RV-variations is {\em larger} than errors
of the measurements. The probability that this is just a coincidence is
0.09. For the rightmost line, it is expected that 0.3\% of the
data-points are right of this line and 99.7\% are left of it. Again we
find that 2 out of 6 measurements are on the line or even right of
it. The probability that this is a chance coincidence is 0.0001. We thus
conclude that the object shows RV-variations.

There are of course two possibilities: One is that this BD is in fact
orbited by a planetary mass object. The other explanation is that the
RV-variations are caused by spots. For stars in the Hyades, which have a
similar age as LP\,944-20, Paulson (\cite{paulson02}) derived a relation
between $v\,sin\,i$ and the rms-scatter of the RV. For a $v\,sin\,i$ of
30 km/s, the rms-scatter of the RV of a star would be 80~m/s. The
RV-variations of LP\,944-20 are thus larger than those of stars of the
same $v\,sin\,i$.  Saar et al. (\cite{saar97}) derived a relation
between $v\,sin\,i$, the filling factor, and the amplitude of the
RV-variations. If we put in 1500\,m/s for the amplitude of the
variations, the filling-factor would be 9\%. Such a large filling-factor
is unusual for stars of the same age and $v\,sin\,i$ as LP\,944-20. A
dark spot covering 9\% of the surface of the star would not be
consistent with the photometric measurements either.  We can also use
our spectra to measure the temperature fluctuations. As pointed out by
Reid et al. (\cite{reid95}), the ratio of the fluxes in the 7042 to 7046
\AA -band versus the flux in the 7126 to 7135 \AA -band can be used for
determining the temperatures of very late-type objects. In the original
paper, this relation is derived only up to the spectral type M5. For
spectral types later then M7, a similar relation can be
derived. LP\,944-20 has a temperature of about 2350\,K which falls 
into the region where this relation between the flux-ratio and the
temperature works well. We derived this ratio for all our spectra of
LP\,944-20, and find that the variations are $\leq 20\pm10$\,K
peak-to-peak. In agreement with previous photometric data, we thus
conclude that there are no big spots on this object.  On the other hand,
one may envision that a BD looks more like a planet than a star. For
example, Saturn has a jet-stream at the equator with a peak velocity of
500~m/s. If part of the jet-stream layers are covered by clouds,
RV-variations would be induced without large variations of the
brightness. We thus conclude that the RV-variations are either caused by
surface features that do not resemble dark sunspots, or by an orbiting
body. With our data it is of course entirely impossible to derive an
orbit for the hypothetical object. If the companion would have a period
of less than two years, its mass would be in the planetary regime 
($\leq 13 M_{\rm J}$).

LHS\,2065 is even more active than LP\,944-20, as large flares
have been observed in X-rays and in the optical on this object (Schmitt
\& Liefke \cite{schmitt02}; Mart\'\i n \& Ardila \cite{martin01a}). The
energy released in one of the X-ray flares exceeds even the largest
solar flares by an order of magnitude. Like in LP\,944-20 the
temperature fluctuations of $26\pm10$~K peak-to-peak are small. With a
distance of 8.5~pc (Monet \cite{monet92}), this object also belongs to
the nearest VLMSs. Despite the similarity to LP\,944-20, we do not find
significant RV-variations. The RV given by Tinney (\cite{tinney98a}) of
8.7 km/s agrees fairly well with the value that we derived of $6.6\pm1.3$
km/s.

\section{Discussion and conclusions}

We have observed 25~BDs and VLMSs at least twice. Two objects turned out
to be spectroscopic binaries, and another object which we took only
one spectrum is also spectroscopic binary. The fraction of spectroscopic
binaries in our sample of BDs and VLMSs thus is $12\pm7$~\%. For
solar-type stars Abt \& Levy \cite{abt76} finds that while 65\% of the
stars are binaries, only 9\% have periods of 100 days or less. Although
our sample is very small, we may conclude that the frequency of
short-period binaries of VLMSs and BDs is roughly the same as that of
solar-type stars.  This result is also in good agreement with the
results of Close et al. (\cite{close02b}) who find the same fraction of
visual binaries (separations of less than 3 AU) for low-mass objects
(spectral types M8 and M9) and higher mass (spectral types M0 to M6)
stars.  The discovery of the spectroscopic BD-BD binary PPL15 (Basri \&
Mart\'\i n \cite{basri99}) and the numerous visual BD-BD binaries also
points in the same direction.

Significant RV-variations where detected in LP\,944-20. These variations
could either be caused by an orbiting planet, or by surface features.
Since LP\,944-20 occasionally shows large flares, surface features
induced by strong magnetic fields are a possibility. On the other hand,
the small amplitude of the temperature variations implies that
RV-variations are not caused by a normal star-spot. Additionally,
LHS\,2065 shows an even larger flare activity than LP\,944-20, but does
not show significant RV-variations.

For the other 19 objects, we do not find significant RV-variations.
While it is impossible to strictly exclude an orbiting planet from
sparsely sampled RV-data, we find that it is unlikely that these objects
are orbited by planets with the mass of a few $M_{\rm J}$, and periods
of 40\,days or less.  This result is interesting because if VLMSs and
BDs have giant planets, these are expected to have periods of
40\,days or less.  While it is certainly not possible to exclude the
presence of massive planets in orbit around VLMSs and BDs, this work is
another piece of evidence that a number of giant planets in orbit around
low-mass objects is equal to or less than that of higher mass stars.

\begin{acknowledgements}

We would like to thank the ESO staff for carrying out the service
observations for all this work. We thank the referee Eduardo
Mart\'\i n for very helpful comments and suggestions for improving the
paper.

\end{acknowledgements}


\begin{thebibliography}{}

\bibitem[1976]{abt76} Abt H.A., Levy, S.G. 1976, ApJS 30, 273

\bibitem[2002]{apai02} 
       Apai, D., Pascucci, I., Henning, Th., Sterzik, M.F., Klein,
       R., Semenov, D., Guenther, E., Stecklum, B. 2002, ApJ 573,  L115

\bibitem[2002]{armitage02} Armitage, P.J., Bonnell, I.A. 2002,
        MNRAS 330, L11

\bibitem[1982]{balthasar82} Balthasar, H., Thiele, U., Woehl, H. 1982,
        A\&A 114, 357

\bibitem[1999]{basri99}Basri, G., Mart\'\i n, E.L. 1999, AJ 118, 2460

\bibitem[2001]{berger01} Berger, E., Ball, S., Becker, K.M., Clarke, M.,
        Frail, D.A., Fukuda, T.A., Hoffman, I.M., Mellon, R., Momjian, E.,
        Murphy, N.W., Teng, S.H., Woodruff, T., Zauderer, B.A., Zavala,
        R.T. 2001, Nature, 410, 338 

\bibitem[1998]{benedict98} Benedict, G.F., McArthur, B., Shelus, P.J.,
        Jefferys, W.H., Duncombe, R.L., Nelan, E.,
        Franz, O.G., Wasserman, L.H., Hemenway, P.D.,
        van Altena, W., Story, D., Whipple, A., Bradley, A.,
        Fredrick, L. W. 1998, Bulletin of the American Astronomical Society, 
        Vol. 30, 1142

\bibitem[1985]{caccin85}Caccin, B., Cavallini, F., Ceppatelli, G., 
        Righini, A., Sambuco, A. M. 1985, A\&A 149, 357 

\bibitem[2000]{cameron00} Comer\'on, F., Neuh\"auser, R., Kaas,
        A.A. 2000, A\&A 359, 269

\bibitem[2000]{chabrier00} Chabrier, G., Brassard, P., Fontaine, G.,
        Saumon, D. 2000, ApJ, 543 , 216

\bibitem[2002a]{close02a} Close, L.M., Potter, D., Brandner, W., Lloyd-Hart, 
        M., Liebert, J., Burrows, A., Siegler, N. 2002, ApJ 566, 1095

\bibitem[2002b]{close02b} Close, L.M., Siegler, N., Potter, D.,
        Brandner, W., Liebert, J. 2002, ApJ, 567, L53

\bibitem[1998]{delfosse98} Delfosse, X., Forveille, T., Mayor, M.,
        Perrier, C., Naef, D., Queloz, D. 1998, A\&A 338, L67

\bibitem[1999]{delfosse99} Delfosse, X., Forveille, T., 
        Beuzit, J.-L., Udry, S., Mayor, M., Perrier, C. 1999,
        A\&A 344, 897-910

\bibitem[1998]{delfosse98} Delfosse, X., Forveille, T., Mayor, M.,
        Perrier, C., Naef, D., Queloz, D. 1998, A\&A 338, L67

\bibitem[1999]{desidera99} Desidera, S. 1999, PASP 111, 1529

\bibitem[2001]{fernandez01} Fern\'andez, M., Comer\'on, F. 2001, A\&A
        380, 264

\bibitem[2002]{gaudi02} Gaudi, B.S., Albrow, M.D., An, J., Beaulieu,
        J.-P., Caldwell, J.A.R., DePoy, D.L., Dominik, M., Gould, A.,
        Greenhill, J., Hill, K., Kane, S., Martin, R., Menzies, J., Naber,
        R.M., Pel, J.-W., Pogge, R.W., Pollard, K.R., Sackett, P.D., Sahu,
        K.C., Vermaak, P., Vreeswijk, P.M., Watson, R., Williams, A. 2002,
        ApJ 566, 463

\bibitem[2002]{goto02} Goto, M., Kobayashi, N., Terada, H., 
        Gaessler, W., Kanzawa, T., Takami, H., Takato, N., Hayano, Y., 
        Kamata, Y., Iye, M., et al. 2002, ApJ 567, L59

\bibitem[2001]{joergens01} Joergens V., Guenther, E. 2001, A\&A 379, L9

\bibitem[2001]{kenworthy01} Kenworthy, M., Hofmann, K.-H., Close, L., 
        Hinz, Ph., Mamajek, E., Schertl, D., Weigelt, G., Angel, R., 
        Balega, Y.Y., Hinz, J., et al. 2001, ApJ 554, L67

\bibitem[2001]{lane01} Lane, B.F., Zapatero Osorio, M.R., Britton, M.C.,
        Mart\'\i n, E.L., Kulkarni, S.R. 2001, ApJ 560, 390L

\bibitem[2000]{lin00} Lin, D.N.C., Papaloizou, J.C.B., Terquem, C., 
        Bryden, G., Ida, S. 2000, in Protostars and
        Planets IV (Book - Tucson: University of Arizona Press; 
        eds. Mannings, V., Boss, A.P., Russell, S.S.), p. 1111

\bibitem[1988]{Lunine88} Stevenson, D.J., Lunine, J.I. 1988,
        Icarus 75, 146

\bibitem[2000]{martin00} Mart\'\i n, E. L., Brandner, W., Bouvier, J., 
        Luhman, K.L., Stauffer, J., Basri, G., Zapatero Osorio, M.R., 
        Barrado y Navascu\'es, D. 2000, ApJ 543, 299
        
\bibitem[2001]{martin01a} Mart\'\i n, E.L., Ardila, D.R. 2001, AJ 121, 2758

\bibitem[2001]{martin01b}Mart\'\i n, E. L., Dougados, C., 
        Magnier, E., M\'enard, F., Magazz\'u, A., 
        Cuillandre, J.-C., Delfosse, X. 2001, ApJ 561, L195

\bibitem[2002]{martin02} Mart\'\i n, E.L., Bouy, H. 2002, 
        New Astronomy 7, 595

\bibitem[1998]{marcy98} Marcy, G.W., Butler, R.P., Vogt, S.S., Fischer,
        D., Lissauer, J.J. 1998, ApJ 505, L147

\bibitem[1992]{monet92} Monet, D.G., Dahn, C.C., Vrba, F.J., Harris,
        H.C., Pier, J.R., Luginbuhl, Ch.B., Ables, H. 1992, AJ, 103, 638

\bibitem[2002]{mullan02} Mullan, D.J., Mac Donald, J. 2001, ApJ 559, 353

\bibitem[2001]{muench01} Muench, A.A., Alves, J., Lada, Ch.J., Lada,
        E.A. 2001, ApJ, 558, L51

\bibitem[2000]{muzerolle00}Muzerolle, J., Brice\~no, C., Calvet, N.,
        Hartmann, L., Hillenbrand, L., Gullbring, E. 2000, ApJ 545, L141

\bibitem[2001]{natta01} Natta, A., Testi, L. 2001, A\&A 376, L22,

\bibitem[2002]{paulson02} Paulson, D.B., Saar, S.H., Cochran, W.D., 
        Hatzes, A.P. 2002, AJ, 124, 572

\bibitem[2000]{persi00} Persi, P., Marenzi, A. R., Olofsson, G., Kaas,
        A. A., Nordh, L., Huldtgren, M., Abergel, A., Andr\'e, P., 
        Bontemps, S., Boulanger, F., Burggdorf, M., Casali, M. M., 
        Cesarsky, C. J., Copet, E.,
        Davies, J., Falgarone, E., Montmerle, T., Perault, M., Prusti, T.,
        Puget, J. L., Sibille, F. 2000, A\&A 357, 219

\bibitem[2002]{potter02} Potter, D., Mart\'\i n, E. L., Cushing, M.C., 
        Baudoz, P., Brandner, W., Guyon, O., Neuh\"auser, R. 2002, 
        ApJ 567, L133

\bibitem[2002]{queloz02}Queloz, D., Santos, N.C., Mayor, M. 2002, In:
        Proceedings of the First Eddington Workshop on Stellar Structure and
        Habitable Planet Finding (eds. F. Favata, I. W. Roxburgh, D. Galadi). 
        ESA SP-485, Noordwijk: ESA Publications Division, ISBN, p.117

\bibitem[1995]{reid95} Reid, I.N., Hawley, S.L., Gizis, J.E. 1995, AJ,
        110, 1838

\bibitem[2001]{reid01} Reid, I.N., Gizis, J.E., Kirkpatrick, J.D., Koerner, 
        D.W. 2001, AJ 121, 489

\bibitem[2002]{reid02}Reid I.N., Kirkpatrick, J. D., Liebert, J.,
        Gizis, J.E., Dahn, C.C., Monet, D.G. 2002, AJ 124, 519

\bibitem[2000]{rutledge00} Rutledge, R. E., Basri, G., Mart\'\i n, E. L., 
        Bildsten, L. 2000, ApJ, 538, L141

\bibitem[1997]{saar97}Saar, S.H., Donahue, R.A. 1997, ApJ, 485, 319

\bibitem[2000]{sasselov00} Sasselov, D.D., Lecar, M. 2000, ApJ, 528, 995

\bibitem[2002]{schmitt02} Schmitt, J.H.M.M., Liefke, C. 2002, 
        A\&A 382, L9
 
\bibitem[2000]{sibille00}Sibille, F. 2000, A\&A 357, 219

\bibitem[1982]{smith82}Smith, M.A. 1982, ApJ 253, 727

\bibitem[1998]{tinney98a} Tinney, C.G. 1998, MNRAS 296, L42

\bibitem[1998]{tinney98b} Tinney, C.G., Reid, I.N. 1998, MNRAS 301, 1031

\bibitem[1999]{tinney99} Tinney, C.G., Tolley, A.J. 1999, MNRAS 304, 119

\bibitem[1998]{trilling98} Trilling, et al. 1998, ApJ 500, 428

\bibitem[2002]{testti02} Testi, L., Natta, A., Oliva, E., D'Antona, F.,
        Comer\'on, F., Baffa, C., Comoretto, G., Gennari 2002, ApJ, 571, L155

\bibitem[2000]{ward00} Ward, W.R., Hahn, J.M. 2000, in Protostars and
        Planets IV (Book - Tucson: University of Arizona Press, 
        eds. Mannings, V., Boss, A.P., Russell, S.S.), p. 1135

\bibitem[1993]{wuchterl93} Wuchterl, G. 1993, Icarus 106, 323

\bibitem[2000]{wuchterl00} Wuchterl, G., Guillot, T., Lissauer, J.J. 2000,
        in Protostars and
        Planets IV (Book - Tucson: University of Arizona Press; 
        eds. Mannings, V., Boss, A.P., Russell, S.S.), p. 1081      

\end{thebibliography}
\end{document}